\documentclass{svmult}

\usepackage{makeidx}     % allows index generation
\usepackage{graphicx}    % standard LaTeX graphics tool
\usepackage{multicol}    % used for the two-column index
\usepackage{epsfig}
\usepackage{amssymb}
\makeindex             % used for the subject index
\begin{document}

\title*{Monte Carlo Studies of Three-Dimensional Bond-Diluted Ferromagnets}
\author{
P.-E. Berche\inst{1},
C. Chatelain\inst{2,3},
B. Berche\inst{2}\and
W. Janke\inst{3}}
\institute{
Groupe de Physique des Mat\'eriaux, Universit\'e de Rouen,
F-76821 Mont Saint-Aignan Cedex, France\\
\textit{pierre.berche@univ-rouen.fr}
\and 
Laboratoire de Physique des Mat\'eriaux,
Universit\'e Henri Poincar\'e, Nancy I,
BP 239,
F-54506 Vand\oe uvre les Nancy Cedex, France\\
\textit{chatelai, berche@lpm.u-nancy.fr}
\and
Institut f\"ur Theoretische Physik, Universit\"at Leipzig,
Augustusplatz 10/11, D-04109 Leipzig, Germany\\
\textit{christophe.chatelain, wolfhard.janke@itp.uni-leipzig.de}}
\maketitle

\section{Introduction}
\label{sec:1}  % Give a unique label
% and use \ref{sec:1} and \cite{journal1}
The influence of quenched, random disorder on phase transitions
is of great importance in a large variety of 
fields \cite{cardybook}, ranging from experiments
with absorbed monolayers \cite{Schwenger94} in condensed matter physics to
conceptual questions in non-perturbative quantum gravity \cite{des}. 

For pure systems exhibiting a {\em continuous\/} phase transition,
Harris \cite{Harris74}
derived the criterion that random disorder is a relevant perturbation when
the critical exponent of the specific heat of the pure system is positive,
$\alpha_{\rm pure} > 0$. In this case one expects that the system falls into 
a new universality class with critical exponents governed by a
``disordered'' fixed point. For $\alpha_{\rm pure} < 0$ the behaviour of the
pure system should persist, and $\alpha_{\rm pure} = 0$ is a special,
marginal case. In two dimensions (2D) this scenario has
been confirmed by various methods for many different systems \cite{selke}.
In three dimensions (3D) extensive
computer simulation studies have concentrated mainly on the 
{\em site\/}-diluted Ising model \cite{selke,Folk2001}.

If a pure system with a {\em first-order\/} phase transition is subject 
to quenched disorder, the transition is softened and may even
turn into a continuous one \cite{ImryWortis79}. This is always the case in
2D \cite{Aizenman89}; for numerical verifications see 
%Refs.~\cite{Chen,Picco,Jacobsen,CC1,Olson}.
Refs.~[9-13].
In higher dimensions, a tricritical point may appear at a finite concentration
of impurities \cite{cardy}, separating ``non-softened'' first-order
and ``softened'' second-order regimes \cite{cardy_statphys}.
Numerically such a scenario has recently 
been observed for the 3D {\em site\/}-diluted 3-state Potts
model \cite{Ballesteros00}. Since the first-order transition
of the pure version of this model is very weak \cite{villanova}, however, the
characterization of the tricritical point remained inconclusive. 

In this report we give an overview on recent results obtained from extensive
Monte Carlo (MC) computer simulations of the 3D 2-state (Ising) \cite{ccp_ising}
%and 4-state Potts \cite{Chatelain01,lat01_potts,ccp_potts} models with 
and 4-state Potts [19-21]  models with 
{\em bond\/}-dilution. 
The motivation to study the 4-state Potts model derives from the fact
that, in the pure case, this model is known to exhibit a fairly strong
first-order transition, such that a disorder-induced softening
to a second-order transition would give clear 
support for the theoretical picture sketched above. Modeling the 
disorder by bond-dilution enables in the Ising case a
test of the expected universality with respect to the type of 
disorder. Furthermore, for both models this choice 
facilitates a quantitative comparison
with recent high-temperature series expansions \cite{meik1,meik2}
for general random-bond $q$-state Potts models.

\vspace*{0.5cm}
\section{Model and Simulation Setup}
\label{sec:2}

The 3D bond-diluted $q$-state Potts model is defined by the Hamiltonian
    \begin{equation}
      -\beta H=\sum_{\langle ij \rangle} K_{ij}\delta_{\sigma_i,\sigma_j};
      \quad \sigma_i=1, \ldots, q,
      \label{eq1}
    \end{equation}
where the sum extends over all pairs of neighbouring sites on a cubic lattice
of size $L^3$ with periodic boundary conditions, and the couplings $K_{ij}$ 
are distributed according to the distribution
    \begin{equation}
      \wp(K_{ij})=p\ \!\delta(K_{ij}-K)+(1-p)\ \!\delta(K_{ij}),
      \label{eq2}
    \end{equation}
where $K \equiv J/k_B T$ is the inverse temperature in natural units.
The dilution parameter $p$ is thus the concentration of magnetic bonds in 
the system,
i.e., $p=1$ corresponds to the pure case.
Below the percolation threshold
$p_c = 0.248\,812\,6(5)$ \cite{Lorenz1998} one does not expect any 
finite-temperature phase transition since without any percolating 
cluster in the system long-range order is impossible.

The model (\ref{eq1}), (\ref{eq2}) was studied by means of large-scale
MC simulations using the Swendsen-Wang (SW) cluster algorithm \cite{SW87} 
in the regime of second-order transitions, and multicanonical 
simulations \cite{muca1,muca2} in the regime where the first-order transition of the
pure 4-state Potts model persists, i.e., at weak dilution close to $p=1$.
Thermodynamic quantities were averaged over a large number of quenched disorder
realisations, ranging between 2\,000 and 5\,000.
The stability of the
disorder averages has been checked by monitoring running averages as
a function of the number of random samples. In fact, some care is necessary
because too small a number of
disorder realisations would lead to {\em typical\/} values rather than
average ones \cite{Derrida84etc}, and these two values are different if
the probability distribution over the disorder realisations exhibits a 
long tail.

\vspace*{0.5cm}
\section{Results}
\label{sec:3}

\subsection{3D Bond-Diluted Ising Model}
\label{sec:3.1}

\begin{figure}
\vspace*{0.2cm}
\epsfysize=5.5cm
\begin{center}
\mbox{\epsfbox{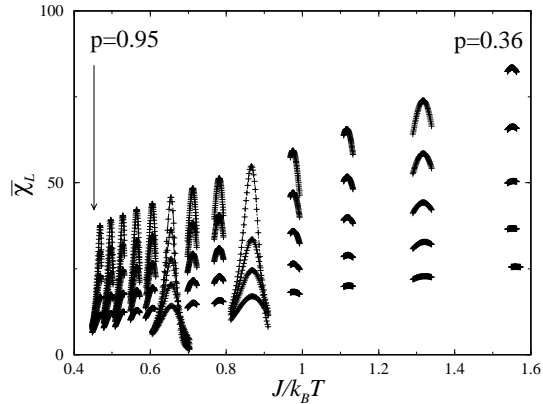}}
\end{center}\vskip -0.3cm
\caption{The average magnetic susceptibility $\bar\chi_L$ of the
3D bond-diluted Ising model versus
$K=J/k_B T$ for several concentrations $p$ and $L=8,10,12,14,16,18$, and $20$.
For each value of $p$ and each lattice size $L$, the curves are obtained by 
standard histogram reweighting of the simulation data at one value of $K$.}
\vspace*{-0.3cm}
\label{Fig1}
\end{figure}

The phase diagram of the 3D bond-diluted Ising model has been
obtained numerically from the locations of the maxima of a diverging quantity
such as the magnetic susceptibility $\bar\chi_L$ depicted in Fig.~\ref{Fig1}.
We focused on $\bar\chi_L$ because the stability of the disordered fixed point
implies a negative specific-heat exponent in a random system \cite{chayes}.
The error in this quantity is hence typically larger than that in the 
susceptibility. 

\begin{figure}[t]
\vspace*{-0.2cm}
\epsfysize=6.3cm
\begin{center}
\mbox{\epsfbox{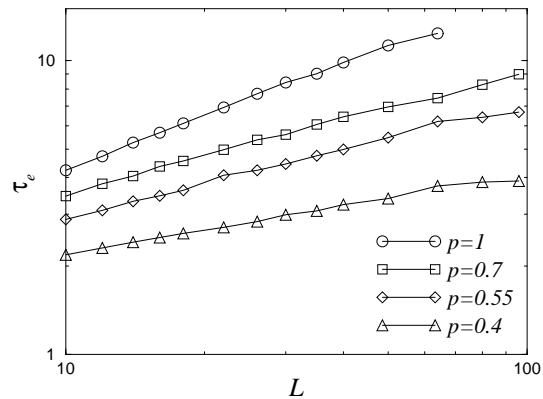}}
\end{center}\vskip -0.5cm
\caption{The energy autocorrelation time $\tau_e$ versus the size $L$ of the
3D bond-diluted Ising model with SW cluster-update dynamics for various
concentrations of magnetic bonds $p$ on a log-log scale. The pure case
corresponds to $p=1$.}
\label{Fig2}
\end{figure}

For an accurate determination of the maxima of the susceptibility we used
the histogram reweighting technique with $N_{\rm MCS} = 2\,500$ MC 
sweeps (MCS) and between $N_s = 2\,500$ and $5\,000$ disorder realisations. The 
choice of $N_{\rm MCS}$ is justified by the increasing behaviour of the energy
autocorrelation time $\tau_e$ as a function of $p$ and $L$. At the critical
point of a second-order phase transition one expects a finite-size scaling
(FSS) behaviour $\tau_e \propto L^z$, where $z$ is the dynamical critical 
exponent. From fits to the data shown in Fig.~\ref{Fig2} we obtained values of 
$z \approx 0.27, 0.38, 0.41$, and $0.59$ for $p = 0.4, 0.55, 0.7$, and 1.0
(the pure case),  respectively. The critical slowing down thus weakens for the 
disordered model and becomes less and less pronounced when the concentration 
of magnetic bonds $p$ decreases. The largest autocorrelation time observed for 
the disordered model was around $\tau_e \approx 9$ for $p=0.7$ and $L=96$.
For each dilution and each size we thus collected at least $250$ effectively 
uncorrelated measurements of the physical quantities, i.e., 
$N_{\rm MCS} > 250\ \tau_e$. 
On the other hand, for small $p \approx p_c$ it is 
necessary to increase the number of disorder realisations because of the 
vicinity of the percolation threshold.

      \begin{table}[t]
       \caption{Evolution of the 3D Ising model susceptibility for $p=0.7$ 
and $L=96$ with the number of MC sweeps, $N_{\rm MCS}$, for different disorder
realisations, $\chi_j$, and the average value over $2\ 500$ realisations,
$\bar \chi$.
        \label{Tab2}}
%\vspace*{0.1cm}
\begin{center}
        \begin{tabular}{|l|l|l|l|l|l|l|} \hline
$N_{\rm MCS}$  & $\ \ \ \chi_1\ $  & $\ \ \ \chi_2\ $ & $\ \ \ \chi_3\ $ &
$\ \ \ \chi_4\ $ & $\ \ \ \chi_5\ $ & $\ \ \ \bar\chi\ $ \\ \hline
$\quad\ 100$ & $\ 1\,268\ $ & $\quad 720\ $ & $\ 1\,141\ $ & $\quad 939\ $ & $\quad 833\
$ & $\ 1\,058\ $     \\
$\quad\ 500$ & $\ 1\,272\ $ & $\ 1\,520\ $ & $\ 1\,223\ $ & $\ 1\,029\ $ & $\quad
953\ $ & $\ 1\,210\ $      \\
$\ \ 1\,000$ & $\ 1\,262\ $ & $\ 1\,544\ $ & $\ 1\,205\ $ & $\ 1\,068\ $ & $\quad
911\ $ & $\ 1\,219\ $     \\
$\ \ 1\,500$ & $\ 1\,282\ $ & $\ 1\,433\ $ & $\ 1\,277\ $ & $\ 1\,047\ $ & $\quad
915\ $ & $\ 1\,227\ $    \\
$\ \ 2\,000$ & $\ 1\,332\ $ & $\ 1\,441\ $ & $\ 1\,221\ $ & $\ 1\,073\ $ & $\quad
917\ $ & $\ 1\,235\ $     \\
$\ \ 2\,500$ & $\ 1\,358\ $ & $\ 1\,484\ $ & $\ 1\,234\ $ & $\ 1\,012\ $ & $\
1\,014\ $ & $\ 1\,234\ $      \\
\hline
\end{tabular}
\end{center}
\end{table}

\begin{figure}[b]
\vspace*{0.2cm}
\epsfysize=5.5cm
\begin{center}
\mbox{\epsfbox{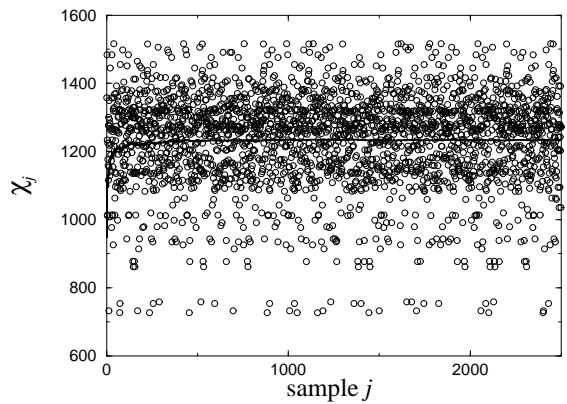}}
\end{center}\vskip -0.3cm
\caption{Distribution of the susceptibility for the different disorder
realisations of the 3D Ising model with a concentration of magnetic bonds $p=0.7$
and $L=96$. The (running) average value over the samples $\bar\chi$ is shown by
the solid line.}
\vspace*{-0.5cm}
\label{Fig3}
\end{figure}

In order to check the quality of the averaging techniques, we first studied
the stability of the susceptibility versus the number of MC sweeps
involved in the thermal average. For the largest size considered
our results are given in Table \ref{Tab2} for different
samples as well as for the disorder average.
With $2\,500$ MCS, the accuracy of the results for a given sample is not
perfect, of course, but the precision of the average over disorder is quite
good on the other hand. The disorder average procedure has been investigated
by computing the susceptibility $\chi_j$ for different samples, $1\leq j\leq
N_s$. As can be seen in
Fig.~\ref{Fig3}, the dispersion of the values of $\chi$ is not very large
because the fluctuations in the (running) average value disappear already 
after a few hundreds of realisations.

\begin{figure}[t]
\vspace*{-0.2cm}
\epsfysize=6.1cm
\begin{center}  
\mbox{\epsfbox{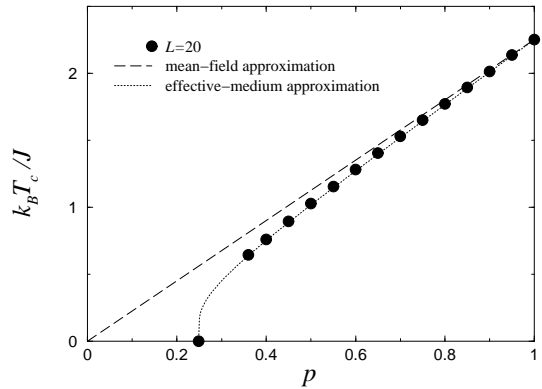}}
\end{center}\vskip -0.5cm
\caption{Phase diagram of the 3D bond-diluted Ising model compared with the
mean-field and effective-medium approximations
(\ref{eq:MF}) and (\ref{eq:EM}), respectively.}
\label{Fig4}
\end{figure}

The phase diagram as obtained from the 
susceptibility maxima for the largest lattice size is shown in Fig.~\ref{Fig4}.
We do not include the results from high-temperature series 
expansions \cite{meik1} in this
figure since they would just fall on top of the MC data. For
comparison we have drawn, however, a simple mean-field (MF) estimate of the 
transition point,
\begin{equation}
K^{\rm MF}_c(p) = K_c(1)/p,
\label{eq:MF}
\end{equation}
with $K_c(1) = 0.443\,308\,8(6)$ \cite{talapov}. This rather crude 
approximation only holds in the low-dilution regime, $p>0.8$. On the other 
hand, the single-bond effective-medium (EM) approximation of 
Turban~\cite{turban}, 
\begin{equation}
K^{\rm EM}_c (p)=\ln \left[(1-p_c ) e^{K_c(1)}-(1-p)\over p-p_c\right],
\label{eq:EM}
\end{equation}
gives very good agreement with the simulated transition line over the full 
dilution range. Since $K_c(1)$ and $p_c$ are input parameters this relation 
is trivially exact in the vicinity of both the pure system
and the percolation threshold.

\begin{figure}[t]
\vspace*{-0.2cm}
\epsfysize=6.cm
\begin{center}
\mbox{\epsfbox{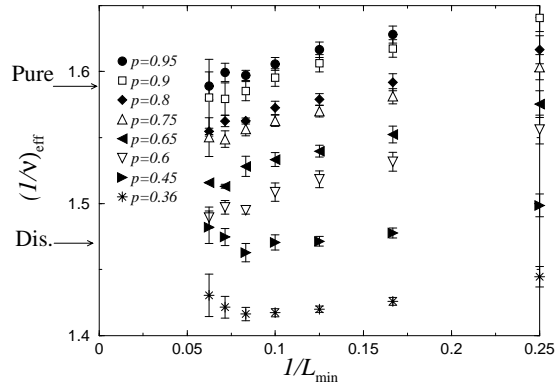}}
\end{center}\vskip -0.3cm
\caption{Effective exponents $(1/\nu)_{{\rm eff}}$ as a function of 
$1/L_{{\rm min}}$ for $p=0.95, 0.9, 0.8, 0.75, 0.65, 0.6, 0.45$, and $0.36$. 
The error bars show the standard deviations of
the power-law fits. The arrows indicate the values of $1/\nu$ for
the pure \cite{Guida1998} and site-diluted \cite{Ballesteros1998} 3D Ising
models, respectively.}
\label{Fig5}
\end{figure}

Due to the competition between different fixed points, one of the main
problems encountered in previous studies of the disordered Ising model was 
the question whether one measures effective or asymptotic exponents. Although
the change of universality class should happen theoretically for an arbitrarily
low disorder, it can be very difficult to measure the new critical exponents 
because the asymptotic behaviour cannot always be reached practically.
Another difficulty comes from the vicinity of the ratios $\gamma/\nu$ and 
$\beta/\nu$ in the pure and disordered universality classes. Indeed, for the
3D Ising model these values are:
\begin{eqnarray}
\!\!\!\!\!\gamma/\nu &=& 1.966(6),\ \beta/\nu=0.517(3),\ \nu=0.6304(13) \quad 
\mbox{pure case \cite{Guida1998}},\\
\!\!\!\!\!\gamma/\nu &=& 1.963(5),\ \beta/\nu=0.519(3),\ \nu=0.6837(53) \quad
\mbox{disordered case \cite{Ballesteros1998}}.
\end{eqnarray}
Recent field theoretical determinations of critical exponents in the disordered 
case are presented in Refs.~\cite{Varnashev2000,Pelissetto2000}, 
and for an excellent review of various experimental, theoretical and numerical
estimates in the last two decades, see Ref.~\cite{Folk2001}.
Thus, from standard FSS techniques, the critical exponent $\nu$ only will 
allow us to discriminate between the two fixed points. This exponent can be 
evaluated from the FSS behaviour of the derivative of the magnetisation 
w.r.t.\ temperature which is expected to behave as 
${d \ln \bar m/ d K} \propto L^{1/\nu}$.  From this power-law behaviour, 
we have extracted the effective size-dependent exponent
$(1/\nu)_{{\rm eff}}$ which is plotted in Fig.~\ref{Fig5} against
$1/L_{{\rm min}}$ for different bond concentrations $p$,
with $L_{{\rm min}}$ denoting the smallest lattice size used in the fits.
We clearly see that in the regime of low dilution ($p$ close to $1$), 
the system is influenced by the pure fixed point. On the other hand, when the 
bond concentration is small, the vicinity of the percolation fixed point 
induces a decrease of $1/\nu$ below its expected disordered value. This is
plausible since the percolation fixed point is characterized by
$1/\nu \approx 1.12$ \cite{Lorenz1998}.

\subsection{3D Bond-Diluted 4-State Potts Model}
\label{sec:3.2}

\begin{figure}[b]
        \epsfysize=5.2cm
        \begin{center}
        \mbox{\epsfbox{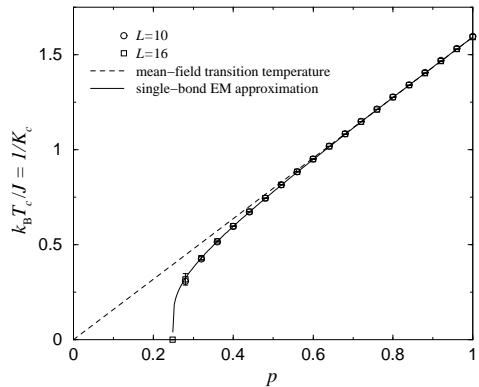}}
        \end{center}\vskip -0.35cm
        \caption{Phase diagram of the 3D bond-diluted 4-state Potts model
        as obtained from MC simulations as well as the
        mean-field (dashed line) and single-bond effective-medium (solid line)
        approximations (\ref{eq:MF}) and (\ref{eq:EM}), respectively.}
        \label{Fig6}
\end{figure}

Let us now turn to the 4-state Potts model which exhibits in the pure case
a rather strong first-order phase transition.
In order to map out the phase diagram of the diluted model
we considered all concentrations $p$ in the interval $[0.28,1]$
in steps of $0.04$ and determined again
the locations of the maxima of the susceptibility
for a given lattice size $L$. The resulting phase diagram is depicted in
Fig.~\ref{Fig6}, where we show for comparison again the simple mean-field
prediction (\ref{eq:MF}) and the effective-medium
approximation (\ref{eq:EM}), using $K_c(1) = 0.62863(2)$ \cite{Chatelain01}.
On the scale of Fig.~\ref{Fig6}, estimates from
high-temperature series expansions up to order 18 are hardly distinguishable,
for a comparison see Ref.~\cite{meik2}.

\begin{figure}[t]
\vspace*{0.3cm}
        \epsfysize=5.2cm
        \begin{center}
        \mbox{\epsfbox{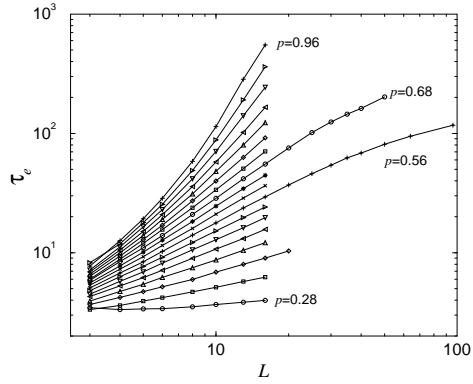}}
        \end{center}\vskip -0.4cm
        \caption{Autocorrelation time $\tau_e$ of the energy at $K_c(p)$
         versus lattice size $L$ ($p$ in steps of 0.04)
         for the 3D bond-diluted 4-state Potts model.}
        \label{Fig7}
\vspace*{-1.3cm}
\end{figure}

In a second step, the order of the phase transitions was investigated. 
Here, in order to satisfy our criterion $N_{\rm MCS} > 250\ \tau_e$,  the 
number of MC sweeps had to be increased up to $15\,000 - 30\,000$,
which is rather large compared to the values used in the Ising case.
A first indication is given by the FSS behaviour of the
autocorrelation time $\tau_e$ at the transition point.
A glance on the log-log plot of Fig.~\ref{Fig7} shows a crossover 
around $p=0.80$ from
a power-law behaviour for strong disorder (small $p$) to a clear exponential 
behaviour for weak disorder ($p \approx 1$), as is 
typical for a first-order phase transition. 
In fact, in the latter case one expects 
$\tau_e \propto \exp(2 \sigma_{od} L^2)$, where the
(reduced) interface tension  $\sigma_{od}$
parameterizes the free-energy
barrier separating the coexisting ordered and disordered phases.

\begin{figure}[b]
        \epsfysize=6.1cm
        \begin{center}
        \mbox{\epsfbox{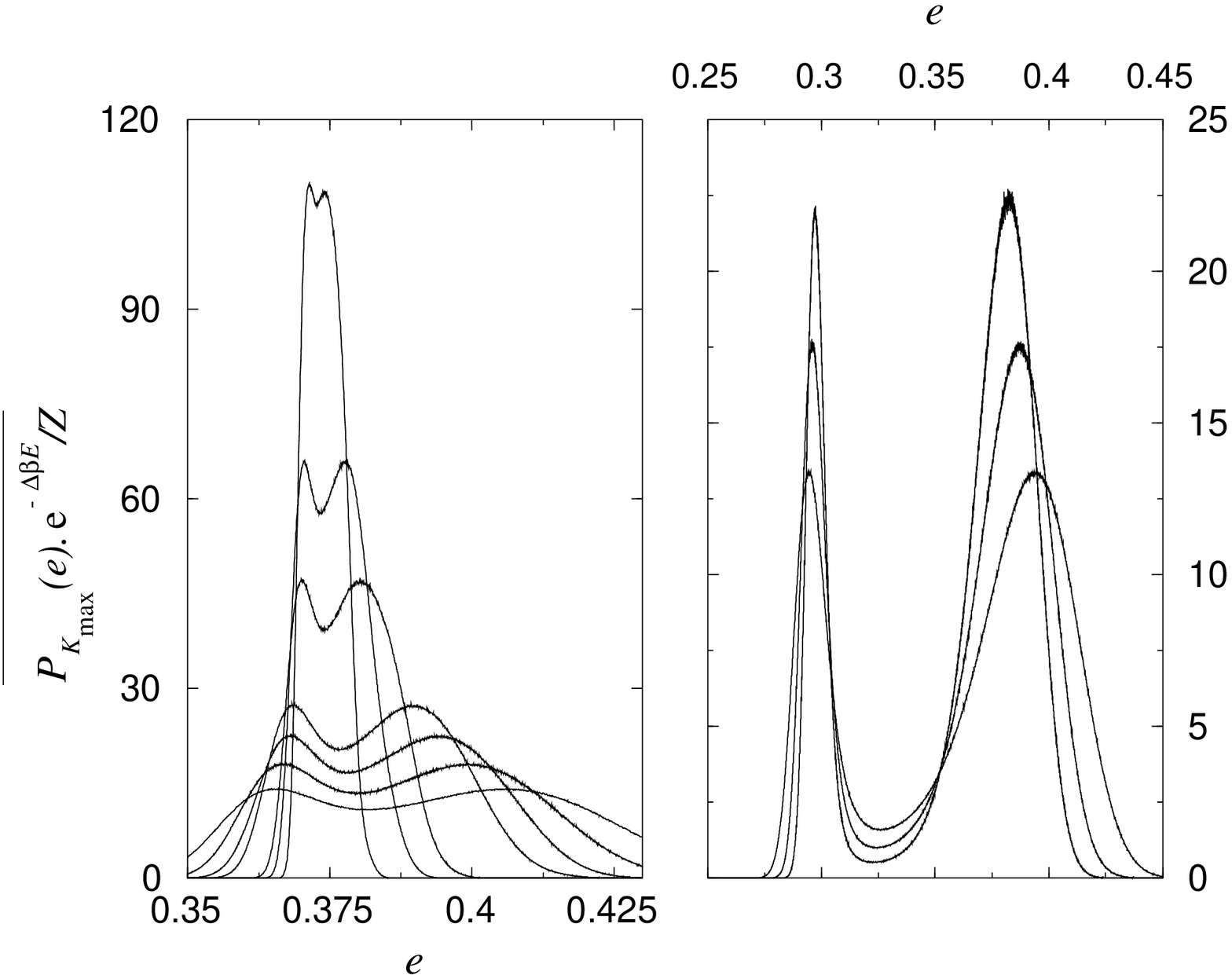}}
        \end{center}\vskip -0.5cm

        \vspace*{-0.1cm}
        \epsfysize=5.0cm
        \begin{center}
        \mbox{\epsfbox{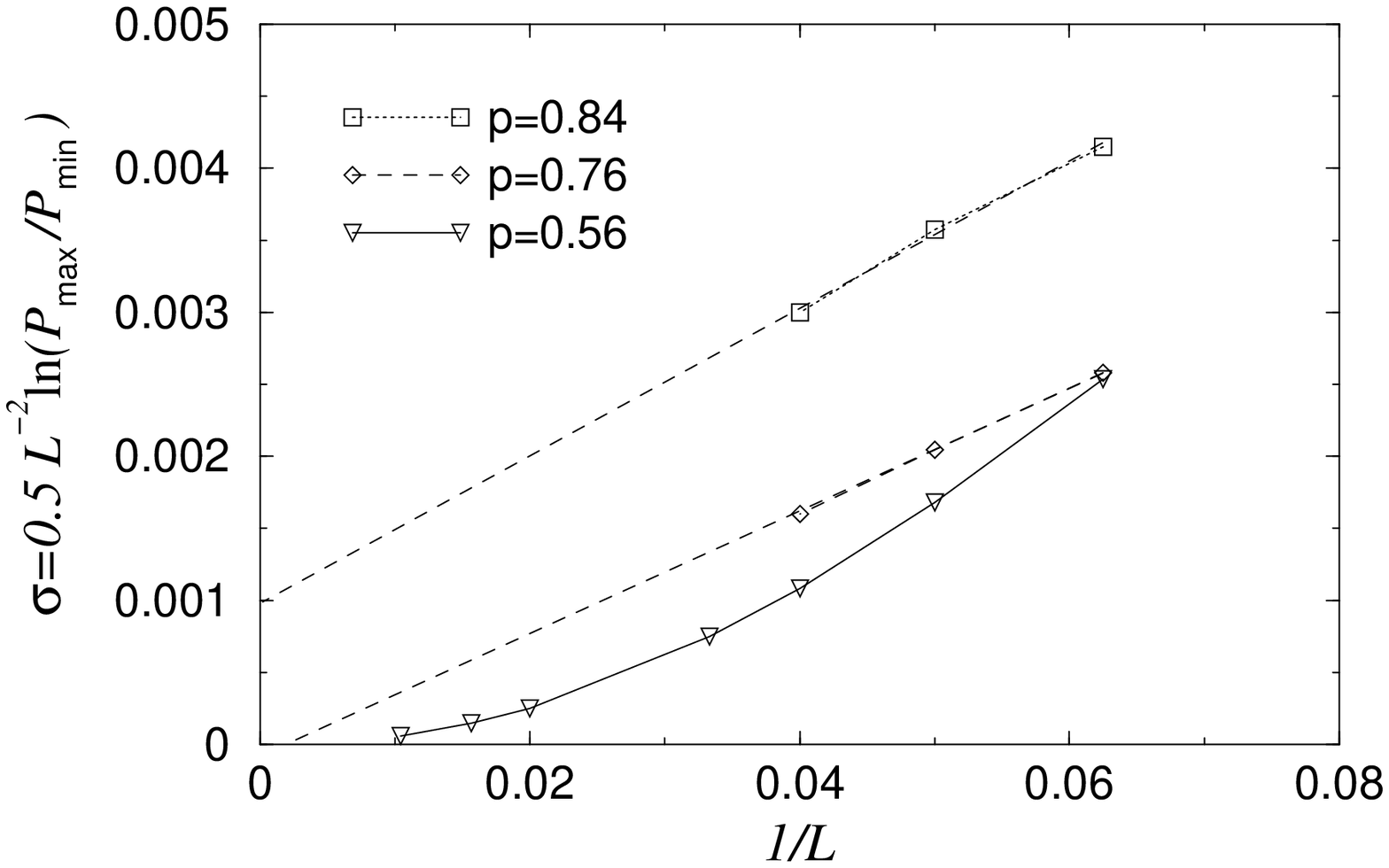}}
        \end{center}\vskip -0.5cm
        \caption{Probability density of the energy of the 
         3D bond-diluted 4-state Potts model reweighted to
          equal peak height for
          $p=0.56$ (top left) and $p=0.84$ (top right).
        Interface tension versus inverse lattice size (bottom).
        \vspace*{-0.1cm}}

        \label{Fig8}
\end{figure}

In the first-order regime we performed multicanonical
simulations and estimated the interface tension from
\begin{equation}
  \sigma_{od} = {1\over 2 L^2} \log{P_{\rm max}\over P_{\rm min}},
  \label{eq4}
\end{equation}
where $P_{\rm max}$ is the maximum of the probability density
reweighted to the temperature where the two peaks are of equal height,
and $P_{\rm min}$ is the minimum in between, see Fig.~\ref{Fig8}.
The linear extrapolations of $\sigma_{od}$
in $1/L$ in the lower part of Fig.~\ref{Fig8} imply
non-vanishing interface tensions only
for $p=0.84$ and above. For $p \le 0.76$, $\sigma_{od}$ seems to vanish
in the infinite-volume limit, being indicative of the expected softening to a
second-order phase transition. The tricritical point would thus be located
around $p=0.76 - 0.84$, in good agreement with the estimate of $p=0.80$ 
derived from the analysis of autocorrelation times.

\begin{figure}[t]
\vspace*{0.4cm}
        \epsfysize=5.2cm
        \begin{center}
        \mbox{\epsfbox{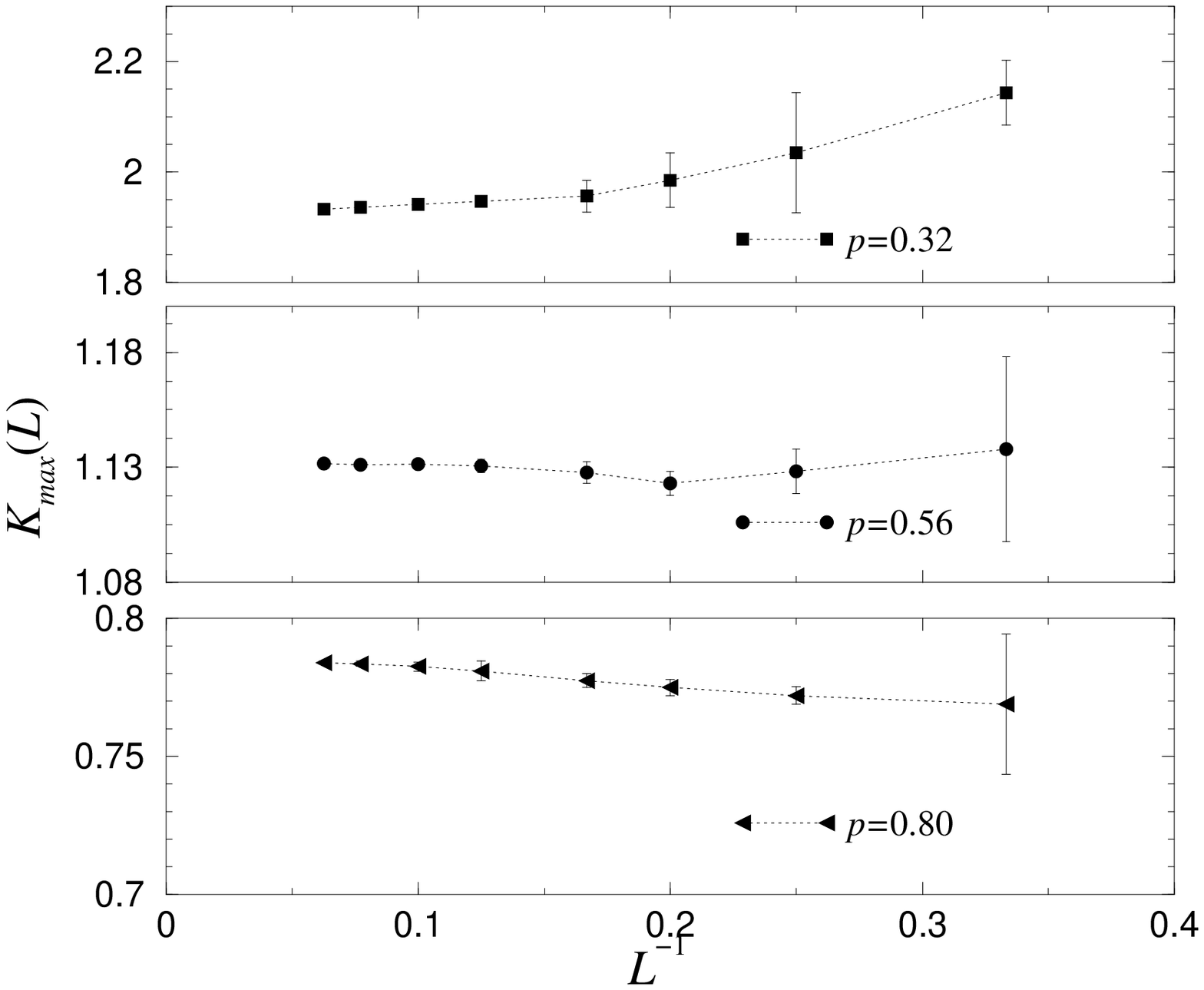}}
        \end{center}\vskip -0.3cm
        \caption{FSS behaviour of the effective transition points for three
        different dilutions of the 3D 4-state Potts model as derived from
        the susceptibility maxima.}
        \label{Fig9}
\end{figure}

To confirm the softening for $p \le 0.76$ we have performed a detailed FSS
study at $p=0.56$ with lattice sizes ranging up to $L=96$ 
and the number of realisations varying between 2\,000 and 5\,000
\cite{Chatelain01}. The choice of $p=0.56$ is motivated by our observation
that in this range of dilutions the corrections to asymptotic FSS of the 
effective transition points are minimal, cf.~Fig.~\ref{Fig9}.

\begin{figure}[t]
\vspace*{0.4cm}
        \epsfysize=5.2cm
        \begin{center}
        \mbox{\epsfbox{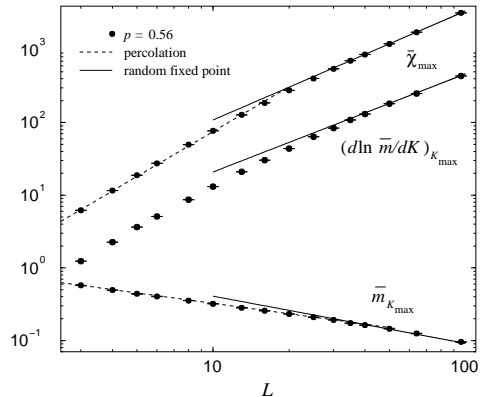}} 
        \end{center}\vskip -0.1cm
        \caption{FSS behaviour of the susceptibility, of
        $d\ln\bar m/dK$, and of the
        magnetization at $K_{\rm max}$ for the
        3D bond-diluted 4-state Potts model at $p=0.56$ (the quantities
        have been shifted in the vertical direction for the sake of clarity).
        The scaling behaviour for small lattice sizes below 
        a crossover length scale is
        presumably governed by the percolation fixed point.}
        \label{Fig10}
\end{figure}

The log-log plot for $\bar\chi_{\rm max}$ in Fig.~\ref{Fig10} shows
that for this quantity the corrections to asymptotic FSS
seem to become quite small
above $L=30$, and fits of the form $a_\chi L^{\gamma/\nu}$
starting at
$L_{\rm min} > 30$ yield $\gamma/\nu = 1.50(2)$.
Using the data for $L < 30$ only, on the other hand, we obtained
perfect fits assuming percolation exponents, 
$\gamma/\nu \approx 2.05$ \cite{Lorenz1998}, cf.\ Fig.~\ref{Fig10}.
Similarly, the FSS of the quantity
$( d \ln \bar m/dK)_{K_{\rm max}} \propto L^{1/\nu}$
gives for $L_{\rm min} > 30$ an estimate of the exponent
$1/\nu=1.33(3)$, consistent with the stability condition
of the random fixed point ($1/\nu \le D/2 = 1.5$) \cite{chayes}.
The same procedure was applied to the magnetization
$\bar m \propto L^{-\beta/\nu}$, but here the
associated critical exponent turned out to be not yet stable.
We therefore also considered the FSS behaviour of higher (thermal) moments of
the magnetization, $\overline{\langle\mu^n\rangle}$, which should scale with an
exponent $n\beta/\nu$. The results for the first moments
exhibit, however, again much stronger corrections to scaling than we
observed for $\bar\chi$ or $ d \ln \bar m/dK$, leading to quite a
conservative final estimate of $\beta/\nu=0.65(5)$.
We nevertheless
note that our results do not fit satisfactorily the scaling
law $2\beta/\nu=d-\gamma/\nu$.
The reason could be the strong corrections to scaling at the random
fixed point which are
hard to cope with for medium-sized systems \cite{Chatelain01}.
%

%\vspace*{0.5cm}
\section{Conclusions}
\label{sec:4}

By performing large-scale Monte Carlo simulations
we have investigated the influence of bond dilution on the critical
properties of the 3D Ising and 4-state Potts models. In the 3D Ising case the 
universality class of the disordered
model is modified by disorder but its precise characterization turned out be
difficult because of the competition between
the different fixed points which induce crossover effects, even for relatively
large lattice sizes.

Applying similar techniques to the 3D
4-state Potts model we obtained clear evidence for softening
to a continuous transition at strong disorder, with estimates for the
critical exponents of $\nu=0.752(14)$, $\gamma=1.13(4)$, and $\beta=0.49(5)$
at $p=0.56$. The analysis of both the
autocorrelation time and the interface tension leads to the conclusion of a
tricritical point around $p=0.80$.

%\vspace*{0.5cm}
\section{Acknowledgements}
\label{sec:5}

We would like to thank Meik Hellmund and Loic Turban for helpful discussions.
The mutual visits within this collaboration were financially supported by the 
joint PROCOPE exchange programme of the DAAD and EGIDE.  C.C. thanks the EU 
network ``EUROGRID: {\sl Discrete Random Geometries: From solid state physics 
to quantum gravity}" for a post-doctoral position in Leipzig, and W.J. thanks 
the German-Israel-Foundation (GIF) for support. The numerical work would have 
been impossible without the computer-time grants 2000007 of the Centre de 
Ressources Informatiques de Haute-Normandie (CRIHAN), hlz061 of NIC, J\"ulich,
and h0611 of LRZ, M\"unchen. We are grateful to all institutions for
their generous support.
%
% BibTeX users please use
% \bibliographystyle{}
% \bibliography{}

%%%%%%%%%%%%%%%%%%%%%%%%%%%%%%%%%%%%%%%%%%%%%%%%%%%%%%%%%%%%%%%%%%%%%%

\printindex
\end{document}